\baselineskip=16pt
\magnification=\magstep1

\newcount\ftnumber
\def\ft#1{\global\advance\ftnumber by 1
          {\baselineskip=12pt \footnote{$^{\the\ftnumber}$}{#1 }}}
\def\section#1{\vskip 12pt \centerline{{\sl #1}} \vskip 6pt}
\font\title = cmr10 scaled 1440        
\font\titlesm = cmr10 scaled 1200        

 \centerline{{\title Making Better Sense of Quantum Mechanics}}
 \vskip 15pt
 \centerline{N. David Mermin}
 
 \centerline{Laboratory of Atomic and Solid State Physics}
 
 \centerline{Cornell University, Ithaca NY 14853-2501}

\vskip 20pt

{\narrower 

\noindent {\it  Abstract.\/}  \hskip 5pt  
We still lack any consensus about what one is actually 
talking about as one uses quantum mechanics.     There is a gap between the abstract terms in which the theory is couched and the phenomena the theory enables each of us to account for so well.   Because it has no  practical consequences  for how we each use quantum mechanics  to deal with physical problems, this cognitive dissonance has managed to coexist with the quantum theory from the very beginning.    The absence of conceptual clarity for almost a century suggests that the problem might lie in some implicit misconceptions about the nature of scientific explanation that are deeply held by virtually all physicists, but are rarely explicitly acknowledged.   I describe here such unvoiced but widely shared assumptions.  Rejecting them clarifies and unifies a range of obscure remarks about quantum mechanics made almost from the beginning by some of the giants of physics, many of whom are  held to be in deep disagreement.  This new view of physics requires physicists to think about science in an unfamiliar way.  My primary purpose is to explain the new perspective and urge that it be taken seriously.   My secondary aims are to explain why this perspective differs significantly from what Bohr, Heisenberg, and Pauli had been saying from the very beginning, and why it is not solipsism, as some have maintained.  To emphasize that this is a general view of science, and not just of quantum mechanics, I  apply it to a long-standing puzzle in classical physics: the apparent inability of  physics to give any meaning to ``Now'' --- the present moment.

}

\vskip 20pt 
{\centerline{\titlesm I.  The stubborn persistence of quantum mysteries.}}
\vskip 10pt

More than ninety years after the invention of quantum mechanics, we find ourselves in a strange situation.   Quantum mechanics works.   Indeed, no theory of physics has ever had such spectacular success.   From ignorance about the structure of matter, quantum mechanics has brought us, in less than a century, to an understanding so broad, powerful, and precise that virtually all contemporary technology relies on it.    And the theory has enabled us to make sense of  phenomena far  beyond anything technology has yet been able to exploit.   

Yet despite this unprecedented success there is notorious disagreement about$\ldots .$ \ \     The sentence fades away because it is not so easy to say what the disagreement actually {\it is\/} about.     Everybody who has learned quantum mechanics agrees  how to {\it use\/} it.    ``Shut up and calculate!''   There is no ambiguity, no confusion, and spectacular success.   What we lack  is any consensus about what one is actually {\it talking about\/} as one uses quantum mechanics.     There is an unprecedented gap between the abstract terms in which the theory is couched and the phenomena the theory enables us so well to account for.    We do not understand  the {\it meaning\/} of this strange conceptual apparatus that each of us uses so effectively to deal with our world.   

 Because it has no bearing on how quantum mechanics  is applied  to  actual physical problems, this confusion has managed to coexist with the quantum theory.  This suggests that the problem might lie in some implicit misconceptions about the nature of scientific explanation deeply held by virtually all physicists, but rarely explicitly acknowledged.   I describe below such unvoiced assumptions that most physicists share.   Acknowledging their existence, and rejecting them as wrong, clarifies and unifies a range of obscure remarks about quantum mechanics by many of the giants of physics, among them people who have traditionally been held to be in deep disagreement.  Abandoning these misconceptions helps make better sense of quantum mechanics.  But it requires physicists to think about science in a radically unfamiliar way.

To date the consequences of what follows are less about how to do physics than about how {\it not\/} to do it.     For decades, some physicists have been searching for modifications in quantum mechanics that lead to changes in its physical predictions too small to have yet been observed.   Such modifications are motivated not by failures of the existing theory, but by philosophical discomfort with one or another of the prevailing interpretations of that theory.  Indicating the false preconceptions at the root of such vexing interpretations could undermine the motivation for such modifications, which  have so far been entirely fruitless.   

What follows may yet have consequences for physics.  But  it has not yet led  to any no significant revisions in how quantum mechanics is used, and I shall not address such possibilities.   
I am not declaring that quantum mechanics will never be modified.  But I am suggesting that if and when quantum mechanics is successfully modified, the motivation will come from unambiguous deviations of actual data from its predictions, and not from discomfort with any interpretations of its formalism. 
\vskip 15pt

The way of thinking about science  I shall describe was developed in the early years of the 21st century by Christopher Fuchs and R\"udiger Schack, collaborating with Carlton Caves until 2006.  Some people maintain that these  developments add little if anything to what Bohr, Heisenberg, and Pauli had been saying from the beginning.   One of my secondary purposes is to make it clear why this is wrong.   Fuchs and Shack are advocating a view of science that is anathema to most physicists.   While they were driven to it in part by the quantum ``no-hidden-variables'' theorems of Bell and of Bell, Kochen, and Specker,\ft{I gave a unified review of the Bell and BKS theorems in N. D. Mermin, Revs. Mod. Phys. {\bf 65} 803-816 (1993), now available, with minor errata incorporated, as arxiv-1805.10311.}  the perspective of Fuchs and Schack also casts an illuminating light on strictly classical physics. 

They have called their understanding of science ``QBism''\ft{More on this terminology in Section 3.4.} (pronounced the same as ``cubism''),  and I have suggested calling it ``CBism'' when the same way of thinking about science is applied to puzzling, if less notoriously vexing, classical issues.

I stress again that what follows need not change how anybody actually uses quantum mechanics.   My purpose is to answer the question of what one is actually doing when one uses it.   Since many physicists  never find it necessary to ask that question, they may well be unpersuaded, perplexed, or even appalled by my answer.  Those who are appalled should bear in mind that I am offering here a philosophical answer --- viewing science as an aspect of the nature of human understanding --- to a philosophical question:  What the hell are we talking about when we use quantum mechanics?   For  {\it practical\/} purposes ordinary everyday quantum mechanics is just fine, and what I have to say is of little or no interest. It is my hope to interest those who, like me, are impractical enough always to have been bothered, at least a bit, by not knowing what they are talking about.

\vskip 15pt

In Section II I set forth the interlocking contents of QBism  as concisely as I can, to encourage
 you to consider them as a whole, before you reject any of them as unacceptable and cease to read others that help to support it.    In Section III I expand my comments on the points mentioned in Section II, and go into their interrelations.  In Section IV I give a QBist view of many different remarks made over the years by scientists, primarily, but not exclusively, physicists.    In Section V, to emphasize that QBism is a view of science in general,  I give the QBist resolution of a long-standing puzzle in strictly classical physics.     
 Section VI is a brief epilogue.

\vskip15pt  
\centerline{\titlesm II. The ingredients of QBism, in brief.}\nobreak\vskip10pt\nobreak\noindent{\it 2.1. Empiricism}\nobreak\vskip5pt\nobreak 
Much of the ambiguity and confusion at the foundations of quantum mechanics stems from an almost universal  refusal to  recognize that {\it individual\/} personal experience is at the foundation of the story {\it each\/} of us tells about the world.  Orthodox (``Copenhagen'') thinking about quantum foundations overlooks this central role of {\it private personal\/}  experience, seeking to replace it by impersonal   
 features of a common ``classical'' external world.
Whenever orthodox thinking addresses the role of experience, it blurs or entirely ignores the difference between one person's experience and another's.\ft{In the case of ``Wigner's friend'' orthodoxy does notice the difference, but takes it to be paradoxical, rather than fundamental.}    

This confusion arises out of a deep desire to objectify and make impersonal what is, at the most fundamental level, unavoidably subjective and personal. To eliminate the confusion it is necessary to acknowledge that science in general, and quantum mechanics in particular, is a tool that {\it each\/} of us uses to organize and make sense of {\it our own\/} private experience.  

I build my understanding of the world I  live in entirely out of my own private personal experience. 
The experiences on which my understanding of my world rests are directly accessible to me and only to me.    
They underlie everything I know about the world.    Like the {\it cogito\/} of Descartes, for me they are self-evident and unquestionable as they happen.\ft{Distinguishing dreams and hallucinations from externally induced experiences is a subtle problem (as it can be in ordinary life) but it is the kind of refinement it would be dis\-tract\-ing to address here.  If I can persuade you that this might be a problem worth further thought, then I will have succeeded in my aim for this essay.}  
A very important component of my personal experience are the words that others address to me,\ft{More on this in point 2.2 immediately below.}
either verbally or in writing, and either individually or as members of collaborating groups or as members of the world-wide community of scientists. 

I have used myself as one example of a perceiving subject.  I could equally well have written  ``you'' or ``Alice'' or ``Bob''.  I  focus on one particular subject, both to underline the unbreakable privacy of the directly perceived experience of each and every one of us, and also to emphasize that  each of us  has their own personal model of the world, based on their own private body of experience.   

What QBism adds to orthodox empiricism, as well as to orthodox interpretations of quantum mechanics, is its insistence that at the most fundamental level ``experience'' does not mean a unique body of common human experience.  Every experience, as I use the term here,  is private to the person having that experience.  So is the picture of the world that each of us constructs from our own experience.

There is indeed a common external world in addition to the many distinct individual personal external worlds. But that common world must be understood at the foundational level to be a mutual construction that all of us have put together from our distinct private experiences,  using our most powerful human invention: language.  

\vskip10pt
\noindent{\it 2.2. Language}

\vskip 5pt

Personal experience is private:  I cannot enter your mind and share your own experience exactly as you perceive it.  It is uniquely and impenetrably yours and only yours.  But language enables each of us to communicate to others crude or even quite sophisticated symbolic representations of our own experience.   By language I mean not only spoken language, but books, scientific publications, emails, gestures, touches, hugs, and so on. 

Because you are part of my external world, like anything else in my world you too can engender experiences in me.  ``Language'' is any way you have to induce experiences in me, through which  you can try  to give me some sense of the content of experiences of your own.  My experience of you provides me with highly plausible evidence that you do indeed have experiences of your own, which play for you the same role that my experience plays for me.   And what you say to me can give me a sense of how you perceive particular experiences of your own.    

My conclusion that you have experiences that are as direct and important for you as mine are for me, is as strong an inference as any I know.   Although it is neither a demonstrable proposition nor a tautology, I would assign it a probability of 1.\ft{More on probability 1 in point 5 below.}  I could not function as a social being were I not certain of this.   I can't believe that you could either.\ft{If you are already  thinking about rejecting all this as solipsism, in spite of my explicit reference to ``you'' and your own experience, please jump ahead immediately to remarks 3.1.2 and 3.3.3.  But then please jump back to here.} 

Language enables us each to hypothesize  features that are common to our otherwise private personal experiences.  This is how we  can each arrive at something like a common understanding of major parts of the worlds that each of us has built from their own experience.    We are in the habit of calling ``the world" this common understanding we all arrive at through language.   Each of us is also all in the potentially more dangerous habit of identifying this less vivid common world with our own private world.   Until people started trying to make sense of quantum mechanics,  this was a useful and usually\ft{But see  Section V.} harmless practice.   

Our 90-year inability to reach agreement on the fundamental meaning of quantum mechanics ought to have brought home to us  the dangers  of each of us  identifying our own personal world with the common world we all negotiate with each other.    Quantum mechanics is a tool that {\it  I\/}  use to help {\it me\/} make sense of {\it my\/} personal world.   It is only 
because language enables us all to conclude that the worlds of different people have features in common, that quantum mechanics can also be a tool that {\it we\/} can 
use to help {\it us\/} make sense of {\it our\/} world.\ft{``We'' here can mean the scientific community as a whole, as well as I and a few collaborators.}   
 
\vskip 10pt
\noindent{\it 2.3.  Actions}
\vskip 5pt  

The world acts on me,  inducing the private experiences out of which I build my understanding of my own world.   Reciprocally, I  act on the world.   I can infer the consequences of my action on the world from the experiences the world creates back in me through its response to my own action.    

My action on the world can be as gentle and implicit as looking at or listening for something, or as intrusive and explicit as building or using or reading about a complex apparatus that performs and records what happens in a carefully designed experiment.   I can ask you a question and experience your answer.    I can perform an experiment and experience the reading of a dial or the output of a computer.    I can listen for the ringing of a bell.   

My actions on the world that lead to a reaction of the world back on me constitute a vast generalization of what are called {\it measurements\/} in orthodox quantum mechanics.     Orthodox measurements are actions mediated by an appropriate apparatus.    The {\it outcome\/}  of a measurement in orthodox quantum mechanics is an objective property of the world: the reading of a pointer, a print-out, a tape, or a computer screen. 

 In QBism the role of measurement is played for me by {\it any action whatever\/} that I  take on the world, whether or not with the help of a ``measurement apparatus''.   The outcome of my action is the {\it experience\/} that the world induces back in me, in response to my action.    My experience of a pointer reading, print-out, return email, etc., are all particular examples of this much more general outcome of a much more general action.      
The outcome of an action, being an experience, is subjective.  It is private to the person taking that action.  But its representation in  language can be communicated to other people.   What I call the objective world is built out of such linguistically shared subjective experiences.

The term ``measurement'' {\it plays no fundamental role\/}  in QBism.   The measurements that play so central a role in the orthodox theory are just particular examples  of actions taken by a user of science,  usually with the help of a large piece of apparatus.    In QBism the {\it outcome\/} of a measurement or  the outcome of any other action by a user, is not a ``classical'' fact as it is in the orthodox theory, but a private experience of that user.   If you are watching me you will have your own private experiences in response to your own action, which is in this case your own observation of my  action.    

Language is the only means we have for trying to compare the  personal outcomes of all such users, and for trying to convey all those outcomes to users who were not watching or otherwise experiencing those actions and their consequences.  

\vskip 10pt
\noindent{\it 2.4. Bettability}
\vskip 5pt

On the basis of my prior experiences I can form expectations for the responses of the world to my actions.   Those  expectations can be  quantified into probabilities.  Those probabilities can be expressed as the odds at which I am willing to place or accept bets.  Those bets should be {\it coherent\/}: I should never be willing to offer and take a group of bets at odds which together result in a certain loss.\ft{Not merely ``in the long run'' but in each individual instance.}   This requirement alone can be shown to imply that my probability assignments must satisfy the usual laws of probability.\ft{See, for example, C. A. Fuchs and R. Schack,  ``Quantum Bayesian Coherence," Reviews of Modern Physics {\bf 85}, 1693-1715 (2013).}    

Laws of science are the regularities we have discerned in our individual experiences, and agreed on as a result of our communications with each other.  Science, in general, and quantum mechanics, in particular,  impose further constraints on my probabilistic expectations. They help each of us place better bets on our subsequent experience, based on our earlier experience.   We are able to navigate the world better because it is bettable.  This is   obvious  for the laws of quantum mechanics, which are explicitly probabilistic.  It also holds for classical physics, though  bettability can be obscured in classical physics by a widespread misunderstanding of probabilities that are often very close to zero or one.\ft{See point 2.5 below.}    

Science can be viewed as a user's guide to the world.   Scientific laws are guides to action, which have proved to be spectacularly successful.   This is an important fact about the world, and about each of our individual worlds.  The laws of science find their meaning in the actions they inspire in every user of science. 

If probabilities are personal expressions of one's willingness to bet,  and one of the laws of quantum mechanics determines probabilities from quantum states, then quantum states must  also be judgments of the person making the state assignment.    Although they describe what one can expect from the world, quantum states are no more objective features of the world than are the probabilities that they encode.

In the orthodox interpretation when a measurement has an outcome the state of the system is said to ``collapse'' to a new state that incorporates that outcome.   In QBism an outcome is  the experience induced in an actor by the world's response to an action.   The collapse of the state is nothing more than the normal updating of the expectations of  that actor on the basis of new experience.  The collapse of a quantum state is conceptually no different from the updating of a probability distribution on the basis of new data.    

I have been chided that quantum states were collapsing in the early universe, before there were physicists.  This makes no more sense than  maintaining that probability distributions were updating in the early universe, before there were statisticians.   What does make sense is for me to reassign a probability to certain events having happened long ago, on the basis of my new current experiences.   It is the same for the collapse of quantum states used to describe phenomena in the early universe.   

 \vskip 10pt
\noindent{\it 2.5. Probability One.}
\vskip 5pt

With the advent of quantum mechanics, deterministic mechanisms disappeared from physics. Should this  be qualified  in a footnote: ``Except when quantum mechanics assigns probability 1 to an outcome"?  
QBism holds that probability assignments are personal judgments even when $p = 1$.  An expectation is assigned probability 1 if it is held as strongly as possible.   Probability 1
indicates a particular intensity of  belief: supreme confidence --- ``I'd bet my life on it.''  It does not imply the existence of a deterministic mechanism guaranteeing the $p = 1$ outcome.  

 An example to keep in mind is my above-mentioned assignment of probability 1 to my belief that you have personal experiences of your own that have for you the same immediate character that my experiences have for me.  This is not a law of nature, backed up by objective properties of the world.  It is my own fundamental guide to action.  

This point was made over 250 years ago by David Hume in his critique of induction. Induction
maintains that if something happens over and over and over again, we can take its
occurrence to be an objective law of nature. What basis do we have for believing in
induction? Only that it has worked over and over and over again.   You can't establish induction without assuming its validity.   Hume concludes that inductive inference is a fundamental human habit.

That probability-1 (or probability 0) assignments are personal judgments, like any other probability
assignments, is essential to the coherence of QBism. This view of probability 1 or 0 has the virtue of undermining the temptation to find any kind of ``nonlocality" in quantum mechanics.   When a single photon is observed at one slit, nothing changes at the other slit when the probability of observing it there instantly drops to 0.  The instantaneous change is in the expectations of whoever made the observation.
 
Most physicists  take it for granted that an outcome with probability 1 must be enforced by an objective mechanism. This was succinctly put by Einstein, Podolsky and Rosen.   Probability-1 judgments, they held, were backed up by ``elements of physical reality''.  Bohr wrote that the mistake of EPR lay in an ``essential ambiguity" in their phrase  ``without in any way disturbing". For a QBist, their mistake is much simpler:  probability-1 assignments, like more general probability-$p$ assignments, are personal expressions of a willingness to place or accept bets, constrained only by the requirement that they should not lead to certain loss in any single event.  When I assign probability 1, I only mean that I'd bet my life on it.

 \vskip 10pt
\noindent{\it 2.6. Object and Subject.}
\vskip 5pt

Physicists' nearly universal conviction, that science should be formulated in a way that makes no reference whatever to the personal experience of the individual user of science, makes it  impossible to express any of the above.   It underlies almost a century  of confusion about the meaning of quantum mechanics.   The fact is that {\it my\/} science has a subject (me) as well as an object (my world).    {\it Your\/} science has a subject (you) as well as an object (your world).    {\it Alice's\/}  science has a subject (she) as well as an object (her world).  I make the same point three times to underline both the plurality of subjects, and the plurality of worlds that each of us constructs on the basis of our own individual experience.   

While each of us constructs a different world,   {\it the\/} world of science is our joint construction of the vast body of phenomena that we try to infer, through language, to be common to our own individual worlds.   Science arises out of our use of language to  indicate to  each other our individual experiences out of which we each construct our own individual worlds.

\vskip15pt

\centerline{\titlesm III. The ingredients of QBism, in more detail.}

\vskip 10pt

\noindent{\it 3.1. Comments on Empiricism}

\vskip 5pt

3.1.1.   Although I am trying to write philosophically about the nature of science, I must acknowledge that  my formal education in philosophy is limited to my first undergraduate year at Harvard, 1952-53, when  the famous Humanities 5 was offered for the first time by Professors Henry Aiken and Morton White.   It was  my great good fortune to be assigned to a discussion section taught by the remarkable Stanley Cavell, then a beginning graduate student. He taught me how to think/write.   I would be pleased if this essay inspired a philosopher  to set my naive  intuitions into philosophically respectable terms.

3.1.2.   Its focus on ``my experience" and  ``my world" has led some  to see QBism as solipsistic.   This is wrong.  I would be foolish not to acknowledge that my experience of {\it you\/} has provided me with overwhelming, if  indirect, evidence that leads me to my unshakeable hypothesis that you have your own private experience that you perceive pretty much as I perceive mine.    From your experience you infer  your own world, which you tell me has for you many of the features that mine has for me.  It is not solipsism if there are many  distinct ``solipsists" comparing notes and trying to construct a common understanding of features they believe to be shared by their own personal worlds.  See also comment 3.3.3 below.

3.1.3.   Direct access to one's own private experience is necessary, but not sufficient for one to be able to use quantum mechanics.   One must also understand the quantum mechanical formalism well enough to be able to deduce revisions of  one's expectations for one's future experience on the basis of one's prior experience.   The answer to one of Einstein's many wonderful skeptical questions is ``No, a mouse cannot collapse a wave-packet'', because no mouse is able to learn and use the quantum formalism to update its expectations. On the other hand a machine also cannot collapse a wave-packet, because although a machine, unlike a mouse, can be programmed to use the quantum formalism,  I see no evidence that a machine has any directly perceived personal experience analogous to mine.  Therefore the actions on the world of a machine have no outcomes for that machine. 

3.1.4.  QBism takes more seriously than do most physicists the basic notion of empiricism, that all knowledge derives from experience. For a QBist  empiricism has a strongly personal flavor to it:   the knowledge of each one of us derives from our {\it own\/} personal experience.    This is close to what William James called ``radical empiricism''.  Different people with different experiences will in general have different knowledge.    The answer to the skeptical question ``whose knowledge?" is ``the knowledge of whoever is using quantum mechanics.''    The answer to ``knowledge about what?'' is ``knowledge about the experience of that knower''.    Collective or joint knowledge is a higher level concept, brought about by efforts to share individual knowledge through the imperfect medium of language.   

3.1.5 Indeed, ``knowledge'' is not the appropriate term for QBist empiricism, in that it suggests something in one-to-one correspondence with impersonal features of the world.  A better term is ``belief". ``Belief" characterizes a model of the world that  particular people construct from their own particular experience. ``Belief" brings into the story a believer. ``Knowledge", insofar as it suggests something more  than a personal inference, does not automatically raise the question of the identity of the knower.

3.1.6 What is most important about scientific belief is that it is constantly subject to challenge and revision (updating) on the basis of new experience, including what we have learned from others about their own beliefs, based on their own experiences.    As I put it elsewhere:\ft{{\it It's about Time: Understanding Einstein's Relativity,\/} Princeton UP, 2005, final paragraph.) .}
 
 {\narrower \narrower
 
\noindent What makes the pursuit of science so engrossing is to learn that one's most strongly held beliefs can be completely wrong. The search to identify and correct the old errors can lead to deep insights into nature. The world would be a far better place for all of us if this joy scientists find in exposing their own misconceptions were more common in other areas of human endeavor.
 
 }

3.1.7.  There is a little remarked upon but important ambiguity in the first person plural.   When Heisenberg says that quantum states are about {\it our\/} knowledge, ``our" can mean all of us collectively or it can mean each of us individually.   Failing to recognize the distinction between these two quite different meanings is responsible for much of the confusion surrounding the ``Copenhagen interpretation''.   When a QBist says  ``Quantum states encapsulate our belief,''  the ``our'' always means each of us individually.   To avoid ambiguity it is better to say  ``My (your, Alice's) quantum state assignments encapsulate my (your, her) belief''  to avoid misreadings based on an implicit assumptions of a unique state assignment  or of common knowledge.  QBism replaces the orthodox term ``knowledge" with ``belief", to sharpen the distinction between individual  belief  based on personal experience, and shared knowledge, common to many people ranging from a few friends through the hundreds of authors of a high-energy physics paper to all of scientifically literate humanity, as a result of the effort to extract what is common to personal belief  through the use of language.

3.1.8.    My own science is at its root entirely about my own personal experience.   The world that each of us infers from our own personal experience is often taken to be what science is about.  Most of the time this is a hair-splitting distinction.   But it is the failure to distinguish between my experience and the world that I infer from my experience that is the root of the confusion at the foundations of quantum mechanics.   Whatever cannot ultimately be traced back to my personal experience cannot be a part of my science.   This need not mean that there are no such entities; just that if they exist, they are beyond the scope of {\it my\/}  science to address.   My experience, crucially,  includes what I hear or read about from others, as expanded on in Section 3.2 below.   This is why there is an enormous common residue shared by my science, your science, and the science of all of us together.  That common residue is generally called simply ``science''.

 3.1.9.  Lurking in most orthodox interpretations of quantum mechanics is a notion of a ``classical'' level of reality.  Landau and Lifshitz\ft{See Section 4.8 below.} are painfully explicit about this.   Sometimes ``classical'' is replaced by ``macroscopic" or ``irreversible".  From a QBist point of view all of these objective, impersonal, vaguely delineated formulations are nothing more than efforts to avoid having to invoke the bedrock of subjective, individually personal  experience.  
 
\vskip 10pt

\noindent{\it 3.2. Comments on Language}

\vskip 5pt

3.2.1.   Niels Bohr always emphasized the importance of ``ordinary language'' because, he said,      
it was the only way to represent to others the outcomes of experiments. Bohr took such outcomes to be objective phenomena in a common external  world, and this unambiguous reality of experimental outcomes has been taken for granted in all subsequent interpretations of quantum mechanics.\ft{The only exception I know of is the many-worlds interpretation, in which all outcomes take place in branches of a ramifying universe.  QBism regards this as the {\it reductio ad absurdum} of reifying the quantum state.}     In QBism, however,  outcomes are  not limited to ``experiments" and they are neither objective nor commonly held.  An outcome is the personal experience induced in a user by the world's response to his or her personal action.  As such it is private and subjective.  

3.2.2. For a QBist,  language --- ordinary or technical --- is  even more essential than it was for Bohr. Language is the only way one can try to represent to others one's  private personal experience, or  get from others a sense of their own private experience.   Science concerns that which is common to the worlds each of us infers from our own unique experience. Language is essential to an understanding of our common science because it is the only way we have to get a sense of what is common to our different private experiences.

3.2.3  Since science is what we can distill out of our individual personal experiences through our efforts to represent those experiences in language, science itself is a form of language --- a very powerful one.   In one sense this seems obvious: we learn science from books and teachers.   Both represent the relevant experiences to us verbally.   In another sense it seems obviously wrong: the laws of physical science can be cast in mathematical forms having a reality that transcends the mere creatures that discover them.    But mathematics itself is a form of language, even more highly refined than science.\ft{Bohr thought so.   See 4.2.5.}     Mathematical Platonists will recoil from this in horror.   The rest of us, however, might think about Wigner's surprise at the ``unreasonable effectiveness'' of mathematics in science.  Why shouldn't mathematics, the most highly refined form of language, serve as the natural vehicle for expressing the most fundamental forms of science, another only slightly less  refined form of language.   Both mathematics and science are human creations.    Even if you are convinced that I should not have said ``creations'', but ``discoveries'', I hope that you will still be willing to consider that what we are comparing here are two representations of those discoveries or creations in human language.  

3.2.4  Without intelligent beings there cannot be language.   Though many mathematicians might disagree, it makes no sense to insist that  the integers would obey the unique factorization theorem even if there were no intelligent beings 
to define the integers and formulate the theorem.  


\vskip 10pt
\noindent{\it 3.3. Comments on Actions}
\vskip 5pt

3.3.1. Fuchs and Schack have called someone who acts on the world an ``agent''.   I prefer the term ``user" [of quantum mechanics].    ``User" emphasizes the fact that quantum mechanics is a tool that anybody who learns it can {\it use\/} to help organize their experiences and anticipate the likelihood of the outcomes of their subsequent actions.  

3.3.2.   The primacy of ``measurement'' in the orthodox interpretation of quantum mechanics is a manifestation of the historic refusal to acknowledge that the user of science has an active role to play in the story.   ``Measurement'' replaces all the different users and their different actions with a single objective material apparatus.  And  ``measurement outcome'' replaces the diverse personal experiences of many users with a single  
reading of a single pointer.  


3.3.3  When I act on the world, the action I take is my free choice.   Reciprocally, when the world acts back on me in response to my action, quantum mechanics can tell me nothing more than the probability of the response I can expect: the world's reaction to my action is the world's free choice.    It would be a strange kind of solipsism that endowed the world with the ability to act on me in ways inherently out of my control.   See also comment 3.1.2, above.


\noindent{\it 3.4. Comments on Bettability}
\vskip 5pt

3.4.1  ``QBism'' originally stood for ``quantum Bayesianism'', because it started off by taking seriously the ``personalist Bayesian'' view that probabilities are subjective personal judgments,  and exploring the consequences of this understanding of probability for our understanding of quantum mechanics.  

3.4.2. But ``Bayesian'' does not capture the view of probability as a personal judgment.  Some Bayesians take a firmly objective view of probability. 

3.4.3.  Nor is it  necessary to become acquainted with technical issues in the foundations of probability theory to acquire a sense of what  QBism is about, as I hope my remarks above have already made clear.  

3.4.4  Most importantly, I would say that developing  an interpretation of quantum mechanics as an outgrowth of a subjective view of probability puts the cart before the horse.  While taking a subjective view of probability does lead unavoidably to the QBist interpretation of quantum mechanics, it is the success of QBism in clarifying the interpretation of quantum mechanics that is by far the strongest argument I know of for taking seriously the view that probabilities --- and even ``laws of nature''\ft{Recall the former Law of Conservation of Mass.}  ---  are ultimately best understood as personal judgments and not objective facts.  
 
3.4.5.  The term ``QBism" nevertheless remains apt, in that  it describes a break with 19th century views of science that is just as great as the break of cubism with 19th century views of art.   Fuchs has pointed out that there already exists an ungainly  but entirely appropriate term beginning with {\it B\/}, {\it bettabilitarianism\/} coined by Oliver Wendell Holmes to describe the analog in epistemology  of utilitarianism in ethics:\ft{The root of ``bettabilitarianism'' (and its most strongly accented syllable) is``bet'' --- what gamblers do.}  

{\narrower\narrower
   
\noindent   I must not say {\it necessary\/} about the universe$\ldots\,.$  We don't know whether  anything is necessary or not.  I believe that we can {\it bet\/} on the behavior of the universe in its contact with us.   So I describe myself as a {\it bet\/}tabilitarian.\ft{{\it Holmes Pollock Letters\/}, Mark DeWolfe Howe (Ed.), Belknap Press, 1961, p. 252.}

}
\vskip 5pt 
\noindent Fuchs has suggested that Holmes'  little-known century-old neologism can accurately  serve for the {\it B\/} in ``QBism'', and I agree.  On the other hand ``QBism'' is acquiring a currency that permits it to stand on its own, without the {\it B\/}  signifying anything at all.  
\vskip 10pt

3.4.6.  To insist that quantum states existed in the early universe, long before anybody existed to assign such states, is to misunderstand the scientific description of the past.     The models of the early universe to which we assign quantum states are models that we construct to account for current experience.  Without current or future data to account for we would have no basis for constructing models of the past.\ft{Rudolf Peierls made this point succinctly in 1991.  See Section 4.10.2.} 

 \vskip 10pt
\noindent{\it 3.5. Comments on Probability One.}
\vskip 5pt

3.5.1.  This may be the hardest aspect of QBism for people  to accept.     If quantum states are subjective judgments, it is absurd to add the proviso that the probability-1 consequences of quantum states are exempt from this status.

3.5.2.   Worse than absurd, it is inconsistent.   Pure states can be {\it defined\/} by specifying a single action having a particular outcome to which quantum mechanics assigns a probability 1.     If that outcome is backed up by an objective ``element of physical reality''\ft{See Section 2.5.}, then that element of reality is a physical mechanism that underlies the state assignment, and the foundations of quantum mechanics plunge back into their 90-year-old fog.  

3.5.3.   I once remarked to R\"udiger Schack that I would never bet my life on anything.   He pointed out  that I do it every time I cross a street.  We build our lives around beliefs we hold with absolute certainty.  Quantum mechanics should teach us to beware of reifying such powerful personal convictions into preexisting elements of reality.  

3.5.4.   David Hume understood this almost two century before quantum mechanics.  The certainties of classical physics are themselves matters of judgment.    The shifty quantum-classical boundary arises from a failure to appreciate this.   Ultimately {\it everything\/} I know rests on my experience.    Beliefs may be held with supreme confidence.   But they remain beliefs.

\vskip 10pt
\noindent{\it 3.6. Comments on Object and Subject.}
\vskip 5pt
3.6.1.  Conventional interpretations of quantum mechanics talk a lot about object vs.~subject.   But the term ``subject" is almost always restricted to the ``observer" or the maker of the measurement.  
Attempts are implicitly made to objectivize the subject.   The references to the ``classical world", or to classical phenomena, or to the measurement apparatus,  or to ``macroscopic" or ``irreversible" phenomena, are all manifestations of the prevailing reluctance  to avoid the subjective --- indeed, the personally subjective ---  aspect of the scientific process. 

3.6.2. Critics of QBism sometimes take it to eliminate the object entirely, in favor of the subject.   This is  a cartoon of a response to the emphasis of QBism that science cannot be understood without reference to the subject.  To be sure, each of us constructs our own objective world on the basis of our own personal experience.   But we then  negotiate among ourselves a common objective world that we can individually and then collectively update and revise on the basis of further individual or verbally shared experience.    

3.6.3.   The objective world asserts itself unmistakably, unpredictably and uncontrollably in its immediate response to any of our interventions.   That immediate response only alters the world of the person or people who experience it, but when the experience induced by the intervention is reported to many, a version of the alteration is transmitted to their common world. 

\vskip 15pt
\centerline{\titlesm IV.  Some quotations that QBism illuminates and vice-versa}  
\vskip 10pt
This Section offers a third opportunity to examine the principles of QBism, this time using them to comment on a collection of quotations.  The quotations are arranged alphabetically by writer and are not intended to develop a single line of thought.    I am not suggesting that any of the writers were early QBists.   What I do want to demonstrate is how QBism can shed light on remarks which may be obscure, paradoxical, or apparently incompatible.    It can also be illuminating to note how a quotation  {\it differs\/} from QBism. 

\vskip 10pt

\noindent 4.1 {\bf John S.~Bell}\ft{Selected Correspondence of Rudolf Peierls, v.2, Sabine Lee [ed], World Scientific, 2009.}

\vskip 10pt  
4.1.1 {\sl Our students learn quantum mechanics the way they learn to ride bicycles (both very valuable accomplishments) without really knowing what
they are doing.}
\vskip 5pt  

\hskip 20pt --- J. S. Bell, letter to R. E. Peierls, 20/8/1980
\vskip 10pt 

4.1.2 {\sl I think we invent concepts, like ``particle''  or ``Professor Peierls", to make
the immediate sense of data more intelligible.}
\vskip 5pt
\hskip 20pt ---    J. S. Bell, letter to R. E. Peierls, 24/2/1983
\vskip 10pt
4.1.3 {\sl I have the impression as I write this, that a moment ago I heard the
bell of the tea trolley. But I am not sure because I was concentrating
on what I was writing$\ldots\,.$ The ideal instantaneous measurements of the
textbooks are not precisely realized anywhere anytime, and more or less
realized, more or less all the time, more or less everywhere.}
\vskip 5pt
\hskip 20pt --- J. S. Bell, letter to R. E. Peierls, 28/1/1981
\vskip 10pt

In 2014, at a conference celebrating the 50th anniversary of Bell's theorem,  I learned that these three quotations, which I had come upon among nearly 2,000 pages of Rudolf Peierls' selected correspondence,  were unfamiliar even to those who, like me, had read and admired almost everything the highly quotable John Bell had written about quantum foundations.

 Quotation 4.1.1 suggests a QBist riddle: Why is quantum mechanics like a bicycle?
Answer: Because, while it is possible to learn how to use either without knowing what you
are doing, it is impossible to make sense of either without taking account of what {\it people\/} actually do with them.  Given a bicycle, an intelligent extraterrestrial who had never encountered a human being would be incapable of grasping the nature of such an  artifact.   Without an awareness of its user, a bicycle makes little sense.   One begins to understand a bicycle only when one learns that it is a tool for a creature with two lower limbs with feet at the ends to fit on the pedals and two upper limbs ending in hands capable of grasping the handle bars.    The very names of these parts of the bicycle make explicit reference to the nature of the user.  The nature of quantum mechanics, like that of a bicycle, cannot fully be understood without explicit reference to its user.   But the nomenclature of quantum mechanics is not as helpful as bicycle nomenclature in making this explicit.   On the contrary, it hides the user from sight.   

Quotation 4.1.2  shows that John Bell was willing to consider concepts, as physical
as a particle or as social as the person to whom he is writing, as human inventions designed to help him make  sense of the data that constitute his experience.  

In Quotation 4.1.3 Bell gives a very concise statement of a view that measurements are not limited to laboratory procedures and that the outcomes of measurements are not limited to readings of devices in the laboratory.  

All of these are fundamental QBist views.   But John Bell was not a protoQBist.   No QBist would share his longing for objectively existing  {\it be\/}ables to replace the measured {\it observ\/}ables of quantum orthodoxy.     {\it Perceive\/}ables might do the job, provided one acknowledged that they were joint manifestations of the world and the perceiver.     But I very much doubt Bell would have accepted this.  

Nevertheless, Bell's remarks to Peierls, and the way in which he criticized Copenhagen, lead me to doubt that he would have rejected QBism quite as superficially as some of his current admirers have done.  In these letters he seems open to a human role in the story, even if not at the most fundamental level.   

\vskip 15pt
\noindent 4.2  {\bf Niels Bohr}
\vskip 10pt

4.2.1  {\sl In our description of nature the
purpose is not to disclose the real essence of the phenomena but only to track down, so far
as it is possible, relations between the manifold aspects of our experience.}\ft{Niels Bohr, 1929. In {\it Atomic Theory and the Description of Nature\/}, Cambridge (1934),
p. 18.}

\vskip 15pt
4.2.2 {\sl Physics is to be regarded not so much as the study
of something a priori given, but as the development of methods for ordering and surveying
human experience.}\ft{Niels Bohr, 1961. In {\it Essays 1958-1962 on Atomic Physics and Human Knowledge\/},
Ox Bow Press, Woodbridge, CT (1987), p. 10.}
\vskip 10pt

These two statements, written more than three decades apart, are remarkably similar.  Both resemble a better known and more sensational remark attributed to Bohr by Aage Petersen after Bohr's death.\ft{See 4.12 below.}
Both quotations state that physics is not so much about phenomena, as it is about our experience of those phenomena.   This sounds like QBism, but there is an important difference.  
The phrases ``{\it our\/} experience'' in 4.2.1  and ``{\it human\/}  experience'' in 4.2.2 fail to distinguish between the unique and immediate personal experience of any single individual, and the collective experience of everybody, negotiated and expressed through the medium of human language.   

The private individuality of human experience, central to QBism, is not, as far as I can tell, a part of Bohr's thinking.  Nor can it be found in any other version of quantum orthodoxy that I am aware of.
My guess is that if required to expand on what he meant by ``experience" Bohr would fall back on impersonal data produced by a macroscopic piece of ``classical'' apparatus.  He would not take it to signify the private personal contents of an individual mind.  

\vskip 10pt

4.2.3  {\sl How do we communicate physical experience at all?$\ldots$ By an experiment, we must understand a situation in which we can tell others what we have done and what we have learned.}\ft{Niels Bohr Public Lecture: Atoms and Human Knowledge, Norman Oklahoma, December 13, 1957.}
\vskip 10pt
4.2.4  {\sl {\it However far the phenomena transcend the scope of classical physical explanation, the account of all evidence must be expressed in classical terms.\/}  The argument is simply that by the word ``experiment'' we refer to a situation where we can tell others what we have done and what we have learned and that, therefore, the account of the experimental arrangement and of the results of observation must be expressed in unambiguous language with suitable application of the terminology of classical physics.}\ft{``Discussion with Einstein on Epistemological Problems in Atomic Physics''} 
\vskip 10pt
4.2.5   {\sl Rather than a separate branch of knowledge, pure mathematics may be considered as a refinement of general language, supplementing it with appropriate tools to represent relations for which ordinary verbal expression is unprecise or cumbersome.}\ft{Niels Bohr, {\it The Philosophical Writings of Niels Bohr\/}, vol.~IV, Jan Faye and Henry J.~Folse (eds.), Ox Bow Press, Woodbridge, CT, 1998.  Page 190.}

\vskip10pt

Ordinary language was enormously important to Bohr.     An essential part of an experiment was reporting it to others.  I  always found this puzzling.   If the outcome of an experiment is an objective classical fact, why is it so important to be able to communicate it to other people in ordinary language?  Why can't they just look for themselves?  

QBism explains this.  If the outcome of my experiment is, like the outcome of any of my actions on the world, just my own personal experience, then it is locked up inside of me.  Only by reporting it in words --- there is no other way to represent personal experience --- can I attempt to communicate it to others.  I also need language to compare my own experience with the verbal representations to me by others of their own private experience.   Taking measurement outcomes to be impersonal features of the world obscures, and, indeed, contradicts this fundamental QBist point.

Quotation 4.2.5, however, comes very close to the QBist/CBist view of mathematics expressed above in point 3.2.3.

\vskip15pt

\noindent 4.3 {\bf Rudolf Carnap}
\vskip 10pt
4.3.1  {\sl  Einstein said that the problem of the Now worried him seriously.  He explained that the experience of the Now means something special for man, something essentially different from the past and the future, but that this important difference does not and cannot occur within physics. That this experience cannot be grasped by science seemed to him a matter of painful but inevitable resignation.}\ft{{\it The Philosophy of Rudolf Carnap,\/} P. A. Schilpp, ed., Open Court, La Salle, IL (1963), p. 37.}

QBism (CBism) accounts for the Now.\ft{See Section V.}  It is a problem only for those who insist on excluding the subject from science.  Physics is not only able to deal with the Now, but it is able to demonstrate that the Now behaves in a way that is fully consistent with  normal human behavior, even when relativistic time dilation is a significant effect.

\vskip 15pt

\noindent4.4  {\bf Albert Einstein}
\vskip 10pt
4.4.1   {\sl The Heisenberg-Bohr tranquilizing philosophy --- or religion? --- is so delicately contrived that, for the time being, it provides a gentle pillow for the true believer from which he cannot very easily be aroused.  So let him lie there.}\ft{Letter to Schr\"odinger, 31 May 1928, {\it Letters on wave mechanics\/},  ed. K. Przibram,  Philosophical Library, New York, 1967, p.~31.}

\vskip 10pt

4.4.2  {\sl At last it came to me that time was suspect!$^{\,}$}\ft{J.~R.~Shankland, Conversations with Albert Einstein, American Journal of Physics, {\bf 31}, 47-57 (1963).  The remark is in a 1950 interview.}

\vskip 10pt

4.4.3   {\sl Space and time are modes in which we think and not conditions in which we live.}\ft{Remark to Paul Ehrenfest, according to Aylesa Forsee, {\it Albert Einstein: Theoretical Physicist\/}, MacMillan, N.Y. (1963), p.~81.} 

\vskip 10pt

I include Einstein's remark 4.4.1  to Schr\"odinger, because they were the two founders who most strongly  rejected the Copenhagen interpretation.   Schr\"odinger, alone among the founders, nevertheless took a strikingly QBist view of the relation of science to the scientist.\ft{See Section 4.13.}  Einstein certainly did not.  

The little known remark 4.4.2, made in a 1950 interview, is my favorite of all Einstein aphorisms. It can be read as conveying a CBist point of view toward time.   But Einstein may have been putting the blame on time itself, rather than on the use of time by the scientist.  

Remark 4.4.3, now a firmly established part of the Einstein canon, is definitely apocryphal.   It seems to come from a popular biography, written shortly after Einstein's death, which cites it as a private remark to Ehrenfest, without offering any supporting evidence.    The sentiment is pure CBism, which makes me even more suspicious of its authenticity.   I would love to think that Einstein said and believed this.   But I find the possibiility hard to reconcile with his views about quantum mechanics.  

\vskip 10pt

\vskip 15pt

\noindent 4.5  {\bf Bruno de Finetti}\ft{{\it Theory of Probability}, Interscience, 1990 (preface).}
\vskip 10pt

4.5.1  {\sl The abandonment of superstitious beliefs about the existence of Phlogiston, the Cosmic Ether, Absolute Space and Time$\ldots\,$, or Fairies and Witches, was an essential step along the road to scientific thinking. Probability too, if regarded as something endowed with some kind of objective existence, is no less a misleading misconception, an illusory attempt to exteriorize or materialize our actual\ft{{\rm The Italian text says {\it vero\/}, which is usually rendered as  ``true'' in English translations of this quotation. But this makes little sense, given what de Finetti is saying.  My guess is that what de Finetti meant here was something more like ``actual''.}} probabilistic beliefs.}

\vskip 10pt

De Finetti, in the 1930s, was one of the earliest  and most eloquent advocates of the 
subjective view of probability, common today among statisticians and inherent in QBism,  but anathema to most physicists.   Physicists generally believe that probability is an objective property of events, baked into their very nature by objective features of the world.  But many statisticians, mathematicians, and philosophers take probability to be a numerical measure of the subjective judgment of the person who assigns it, indicating, for example, the odds at which that person is willing to make or accept bets.  

Since probabilities play an irreducibly fundamental role in quantum mechanics, one's understanding of quantum mechanics should differ dramatically depending on whether one subscribes to the objective (``frequentist'') view of most physicists, or the subjective (``personalist'') view of most statisticians.   It is odd that so little attention has been paid to these strikingly different understandings of probability during the ninety years of confusion over the meaning of quantum mechanics.  Quantum mechanics is, after all, the first physical theory 
 in which probability is explicitly not a way of dealing with ignorance of the precise values of existing quantities.   One would have thought that the Born interpretation called for a searching reexamination of the nature of probabilistic judgments.   But until the advent of QBism, in the early 21st century, there was no such inquiry.  

If one does take a subjective view of probability, then a QBist understanding of quantum mechanics is unavoidable.   But I prefer to put it the other way around: the success of QBism in clarifying the murk at the foundations of quantum mechanics is a compelling reason for physicists too to embrace the widespread view of probabilities as subjective personal judgments .

\vskip 10pt  

\noindent4.6 {\bf Sigmund Freud\/}

\vskip 10pt  {\sl A world constitution that takes
no account of the mental apparatus by which we perceive it is an empty abstraction.}\ft{{\it The Future of an Illusion\/}, 1927, concluding paragraph.}

Freud was writing about religion.    But for a QBist his remark applies equally well to physical science, contrary to the current views of most physical scientists.

\vskip 15pt

\noindent 4.7  {\bf Werner Heisenberg}
\vskip 10pt

4.7.1 {\sl The conception of the objective reality of the elementary particles has thus evaporated in a curious way, not into the fog of some new, obscure, or not yet understood reality concept, but into the transparent clarity of a mathematics that represents no longer the behavior of the elementary particles but rather our knowledge of this behavior.}\ft{Werner Heisenberg, Daedelus {\bf 87} (Summer 1958), p.~100.}
\vskip 5pt

This is an example of the kind of quotation by a Founder that leads people superficially to declare that QBism brings nothing new to the story.  Heisenberg's remark certainly has a QBist flavor to it, but it raises questions that nobody, before Fuchs and Schack, has been able to answer.  What does it mean to distinguish between ``behavior'' and ``our knowledge of this behavior''?  To whom does ``our'' refer?  What does it mean to ``know'' something?   What does it mean to confer ``reality'' on something?

Whether or not you like the answers QBism provides, I hope you will agree that they are simple and unambiguous.  As discussed above,  in a QBist reading ``our" does not refer, as it usually does, to all of us collectively.  It refers to each one of us individually.   Some of us may have such knowledge; others may not.   The object of my knowledge is the content of my private personal experience, or, less directly and more inferentially, the world that experience has led me to hypothesize.  ``Know'' may well apply to my experience, but it can be misleadingly strong to apply the word to what I hypothesize.  

\vskip 15pt

\noindent 4.8  {\bf L.~D.~Landau and E.~M.~Lifshitz}\ft{{\it Quantum Mechanics\/}, Pergamon Press,  3rd edition, 1976, pps.2,3.}
\vskip 10pt

4.8.1 {\sl It is in principle impossible
$\ldots$to formulate the basic concepts
of quantum mechanics without using classical mechanics.}
\vskip 10pt

4.8.2 {\sl It must be most decidedly emphasized that we are here not
discussing a process of measurement in which the physicist-observer
takes part.  By measurement, in quantum mechanics, we understand any process of interaction between classical and quantum objects, occurring apart from and independently of any observer.\/}
\vskip 15pt

The quantum mechanics text of Landau and Lifshitz provides the most extreme and explicit example of the broad reluctance of physicists to allow any trace of a human presence to play a role in our understanding of quantum mechanics.\ft{It was suggested to me by Czech colleagues that such quotations are nothing more than Landau and Lifshitz paying lip  service to the Marxist doctrine of dialectical materialism.   I'm unconvinced.   Neither author was ever a model of orthodox Soviet behavior,  and nowhere else in their voluminous writings am I aware of any other tipping of the hat to communist ideology.   I doubt they would have used so strong a formulation in their exposition of basic quantum mechanics if they didn't believe it.}    While other versions of quantum orthodoxy locate measurement outcomes in a vaguely defined ``classical'' world coupled to a quantum mechanical system,\ft{Other common and equally imprecise euphemisms for individual experience are ``ma\-croscopic'' or ``irreversible''.} Landau and Lifshitz are the only ones explicitly to insist that classical mechanics is therefore logically prior to quantum mechanics.   They literally distinguish between ``classical objects'' and ``quantum objects''.  

QBism makes no such distinction.  
The term ``classical'' plays no fundamental role at all in QBism. In orthodox interpretations the term ``classical" enables one to avoid any reference to the directly perceived personal experience of each individual user of quantum mechanics. From the QBist perspective, calling such personal experience ``classical'' is part of the unacknowledged need to depersonalize what is specific to each user of quantum mechanics, and to objectify what is inherently subjective.   

For a QBist there are no ``classical objects'' as opposed to ``quantum objects''.   The roles of classical objects are played by  the personal experiences on which an understanding of the world external to the QBist rests.  The roles of quantum objects are played by the entities that the QBist hypothesizes to comprise that world on the basis of that experience.    

So the QBist reformulation of quotation 4.8.1 from Landau and Lifhistz would be this:  It is impossible to formulate the basic concepts of quantum mechanics without referring to the personal experience of each user.    And the second sentence of quotation 4.8.2 would become this: By measurement, in quantum mechanics, we understand any interaction between a user of quantum mechanics and the world external to that user that results in the world inducing an experience back in that user. 

\vskip 15pt

\noindent 4.9 {\bf Wolfgang Pauli}\ft{W. Pauli, ``Probability and Physics'',  Dialectica {\bf 8}, 112-124 (1954). Translated in {\it W. Pauli, Writings on Physics and Philosophy\/}, edited by C. P. Enz and K. von Meyenn, and translated by
R. Schlapp (Springer-Verlag, Berlin, 1994), pp. 43-48.}
\vskip 5pt

4.9.1 {\sl Nevertheless, there remains still in the new kind of theory an objective reality, inasmuch
as these theories deny any possibility for the observer to influence the results
of a measurement, once the experimental arrangement is chosen. Therefore particular
qualities of an individual observer do not enter the conceptual framework of the theory.}

In QBist terminology:   

Objective reality is maintained in quantum mechanics through the denial of any possibility for a user to influence the particular experience induced by the reaction of the world, once the user's particular action on the world has been chosen.   In specifying the nature and the likelihood of the possible experiences induced by a given action, quantum mechanics makes no reference to the particular qualities of an individual user other than that user's choice of state assignment.

The only clash here with QBism --- and it is a big one --- is Pauli's narrowing of the action of the user into ``the experimental arrangement", and his broadening of the particularity of the user into an unspecified ``observer''.   To be sure, QBism agrees with Pauli that the particular qualities of the user are not a part of the theory, but the fact that each user {\it is\/} particular, plays no role for Pauli, though it lies at the heart of QBism.

\vskip 15pt

\noindent  4.10 {\bf Rudolf Peierls}
\vskip 10pt
4.10.1 {\sl In my view, a
description of the laws of physics consists in giving us a set of correlations between successive
observations. By observations I mean$\ldots$what our senses can experience. That we
have senses and can experience such sensations is an empirical fact, which has not been
deduced (and in my opinion cannot be deduced) from current physics.}\ft{{\it Selected Correspondence of Rudolf Peierls,\/} vol. 2, Sabine
Lee [ed], World Scientific, 2009, p. 807.}
\vskip 5pt

4.10.2 {\sl If there is a part of the Universe, or a period in its history, which is not capable of influencing present-day events
directly or indirectly, then indeed there would be no sense in applying quantum mechanics to it.}\ft{{\it Physics World\/}, January 1991, pps.~19-20.}

\vskip 10pt

Peierls' view 4.10.1 of the laws of physics comes as close to QBism as any statement I can find among the Founders or the earliest post-quantum generation.  If he had taken care to specify that when he said ``we'', ``us'', and ``our'' he meant each one of us, acting and responding as a user of quantum mechanics, he would have had it exactly right.   But it seems to me likely that he was using the first person plural collectively, to mean all of us together, thereby promulgating the Copenhagen confusion.    Contrary to QBism, Peierls also takes it for granted,\ft{Rudolf Peierls, {\it More Surprises in Theoretical Physics\/} Princeton University Press, Princeton, NJ, 1991, p. 11.} like most physicists,  that outcomes that are assigned probability 1 must be backed up by an explicit physically describable mechanism .

In quotation 4.10.2 from his {\it Physics World\/} essay, Peierls articulates 
the QBist view of why it makes sense to apply quantum mechanics to eras long before people existed.    Events in the remote past can influence our present  or  future experiences.


Peierls also refers in that essay to ``the view of Landau and Lifshitz (and therefore of Bohr)''.   This identification of Landau and Lifshitz with Bohr has to be taken seriously, since Peierls and Landau worked together in Copenhagen in the early days.   But I have never read in Bohr anything as literally dualistic as the remarks 4.8.2 of Landau and Lifshitz on classical and quantum ``objects''.  Nor have I read in Bohr anything as close to QBism as Peierls' 4.10.1,  which I would take (perhaps wrongly) to be his own reading of Bohr.

\vskip 15pt

\noindent 4.11 {\bf Asher Peres}
\vskip 10pt

4.11.1 {\sl Unperformed experiments have no results.}\ft{American Journal of Physics {\bf 46}, 745 (1978).}
\vskip 10pt

Peres' famous and charming title, the most concise insight into the nature of quantum mechanics I know of, is reduced by QBism to a truism: Nobody can experience the world's response to an intervention that was never made.  Unexperienced experiences are not experienced.  

\vskip 15pt

\noindent4.12  {\bf Aage Petersen}\ft{A. Petersen, Bulletin of the Atomic Scientists {\bf 19}, 7, 8-14 (1963).}
\vskip 10pt

4.12.1 {\sl When asked whether the algorithm  of quantum mechanics could be considered as somehow
mirroring an underlying quantum world, Bohr would answer ``There is no quantum world.
There is only an abstract quantum physical description. It is wrong to think that the task of
physics is to find out how nature is. Physics concerns what we can say about nature.''}

\vskip 10pt

This, perhaps the most famous of all Bohr quotations about quantum mechanics, is reported in a memorial essay by his amanuensis, Aage Petersen.    Petersen, I am told, was very close to Bohr and knew his views very well.   Although Bohr never published anything quite as strong as Petersen's quotation, his writings cited above have a similar flavor.      Except for the usual ambiguous ``we'' Petersen could have been quoting a QBist.   But the QBist would have said ``each of us'', and expanded on the term ``world''.

\vskip 15pt

\noindent 4.13 {\bf Erwin Schr\"odinger}
\vskip 10pt

4.13.1  {\sl The scientist subconsciously, almost inadvertently simplifies his problem
of understanding Nature by disregarding or cutting out of the picture to be constructed,
himself, his own personality, the subject of cognizance.}\ft{{\it Nature and the Greeks, Science and Humanism\/}, Cambridge (1996), p. 92. See also {\it Mind and Matter\/}  and {\it My View of the World.\/}}

\vskip 5pt
4.13.2 {\sl Quantum mechanics forbids statements about what really exists --- statements about the object. It deals only with the object-subject
relation. Although this holds, after all, for any description of nature, it appears to hold in a much more radical and far-reaching sense in quantum mechanics.}\ft{Schr\"odinger to Sommerfeld, 11 December, 1931, in {\it Schr\"odingers Briefwechsel zur Wellenmechanik und zum Katzenparadoxon\/}, Springer Verlag, 2011.  English translation by N. David Mermin and R\"udiger Schack.}

\vskip 5pt
{\sl   
4.13.3 The conception of a world that really exists is based on there being a far-reaching common experience of many individuals, in fact of all individuals who come into the same or a similar situation with respect to the object concerned.  Perhaps instead of ``common experience'' one should say ``experiences that can be transformed into each other in a simple way''.$\,\ldots$ [This reality] comes about for us as, so to speak, the intersection pattern of the determinations of many --- indeed of all conceivable --- individual observers.   It is a condensation of their findings for economy of thought which would fall apart without any connections$\ldots\,$.   
}\ft{Letter to Einstein, 18 November 1950,  {\it Letters on wave mechanics\/},  ed. K. Przibram,  Philosophical Library, New York, 1967, p.~37.}

\vskip 10pt

 In the 1960s Schr\"odinger repeatedly expressed the view that leaving the scientist out of the story, an exclusion going back to the ancient Greeks, though crucial for the subsequent development of science, nevertheless got things wrong.   He rarely mentions quantum mechanics in this setting.  But thirty years earlier, in a letter to Arnold Sommerfeld, he does explicitly tie this view to quantum mechanics.  Even then, he indicates that it applies to science much more broadly.     Quotations 4.13.1 and 4.13.2 are pure QBism.   Those who maintain that QBism is nothing more than an elaborated version of the Copenhagen interpretation should note that Schr\"odinger shared with Einstein\ft{See 4.4.1.} a scorn for ``the Bohr-Heisenberg tranquilizing philosophy.''  

In his 1950 letter to Einstein, 4.13.3,  Schr\"odinger comes close to articulating a view of {\it the\/} world as the common residue of the experience of many.    While he does not mention the important role of language in such sharing of experience, 
his expansion of the phrase ``common experience'' into ``experiences that can be transformed into each other in a simple way'' indicates his awareness that there is an issue here in need of elaboration.   On the other hand his oft-stated endorsement of  the view that all minds are one, offers quite a different answer.   
\vskip 15pt

\noindent 4.14 {\bf Steven Weinberg}\ft{{\it New York Review of Books\/}, January 19, 2017.  See also Letters, April 6, 2017.}
\vskip 10pt


4.14.1 {\sl It is a bad sign that those physicists today who are most comfortable with quantum mechanics do not agree with one another about what it all means.}

\vskip 5pt

4.14.2 {\sl  
 I hoped for a physical theory that would allow us to deduce what happens when people make measurements from impersonal laws that apply to everything, without giving any special status to people in these laws.

\vskip 5pt

4.14.3 It is not that we object to thinking about humans. Rather, we 
 want to understand the relation of humans to nature, not just assuming the character of this relation by incorporating it in what we suppose are nature's fundamental laws, but rather by deduction from laws that make no explicit reference to humans. We may in the end have to give up this goal, but I think not yet.}

\vskip 5pt

4.14.4 {\sl There seems no way to locate the boundary between the realms in which, according to Bohr, quantum mechanics does or does not apply.}

\vskip 10pt

In quotation 4.14.1 Steven Weinberg shares my concern that the lack of agreement about the meaning of quantum mechanics is a warning that ought to be taken seriously.    QBists  would say that the root of the confusion that worries him lies precisely in his next two quotations, 4.14.2 and 4.14.3, which express the very common\ft{Freud and Schr\"odinger are rare exceptions.} insistence that laws of nature should make no reference whatever to the people who have formulated those laws.   

But why should scientific laws never, under any circumstances, mention any user of science?  Science is a human activity.   The laws are formulated in human language.   As empiricists most scientists believe that their understanding of the world is based on their own personal experience.{\ft {Including  words of others they have heard and read.}   Why insist that  an understanding of science, which I use to make sense of the world I infer from my experience, should make no mention whatever about the role of that experience? 

If one is allowed to let users of science into the story, then there is no problem at all in locating Bohr's boundary in Weinberg's 4.14.4.   The  boundary --- John Bell's ``shifty split''\ft{John S. Bell, ``Against Measurement'', Physics World, August, 1990.}  ---   is different for each scientist using quantum mechanics.  But for {\it each\/} user the boundary is entirely unambiguous: I apply quantum mechanics to the world I construct from my own experience; the role of Bohr's classical world is played {\it for me\/} by that experience itself.

This is quite a modest incorporation of the relation between humans and nature 
into nature's fundamental laws (Weinberg, 4.14.3),   The laws continue to have a universal form, independent of who is using them.   It's just that the domain of application of the laws  ---  the relation between nature and the experience of each particular person applying the laws --- unavoidably varies from one user to another, because the directly perceived personal experience of each person is strictly private to that person.  

QBism  incorporates into nature's fundamental laws nothing beyond an acknowledgment that those laws are stated in human language and rest upon human experience.   This in no way puts the subjects of linguistics or psychology outside the realm of scientific inquiry.    It {\it may\/} suggest that these subjects are harder to disentangle from the common understanding of the material world that physicists strive to construct,  but whether or not aspects of them are intrinsically beyond the scope of physics to address is, as Peierls remarks in 4.10.1, still very much an open question.

\vskip 15pt
\centerline{\titlesm V.  There is no classical world}
\nobreak\vskip15pt

What underlies my world at the bottom is only my experience.  Experience is neither quantum nor classical. Experience is {\it sui generis\/}.    Orthodox interpretations of quantum mechanics, if they mention experience at all, as Bohr often does, depersonalize and/or reify it.   This is why the notion of classicality plays so big a role in orthodox interpretations.   The ``classical" (or the ``macroscopic'' or the ``irreversible") is a catch basin.   Its actual purpose is only to make physicists less uncomfortable, by hiding the subjective and the personal behind something taken to be objective and impersonal.\ft{Particularly the phrase ``classical'' can often be taken to be nothing more than an orthodox euphemism for ``experience''.}  
The many ``paradoxes'' of orthodox quantum mechanics --- for example the lack of any physical mechanism underlying the collapse of a quantum state --- are a sign that this stratagem does not work.        

It once made sense to exclude the scientist from scientific explanations of the physical world.   This warded off  superstitious, animistic, or religious explanations.  But without  endorsing superstition, animism, or religion, today it makes sense to insist that the scientist should {\it not\/} be excluded from a philosophical understanding of the nature of scientific explanation.   Why shouldn't such an understanding involve the explainer, as well as the explained?  It is our exclusion of each individual subject from the story we tell about science that underlies our ninety year failure to agree upon the meaning of quantum mechanics.  

This exclusion of the subject  has sown confusion even in strictly classical physics.    When used in a setting well described by classical physics, QBism  may be called CBism.\ft{N. David Mermin,  Physics Today, Commentary, March 2014; Nature {\bf 507}, 421-3, 27 March, 2014.}  Carnap's report 4.3.1 of Einstein's insistence that there was no room for the {\it Now\/} in physics  is a spectacular example of a puzzle in classical physics resolved by CBism.    When viewed from this perspective, Carnap's story illustrates the mindset that kept Einstein from accepting quantum mechanics. It is worth expanding on.

The experience of the Now does indeed mean something special for me, something essentially different from my past and my future.   The apparent absence from physics of this important difference is an artifact of the unwarranted removal of the subject from the story physics is allowed to tell.   That the Now appears to be unavoidably missing is a clear indication that the world indeed does not make sense,  if I insist on leaving my own experience out of the story I tell about it.   

The special character of my Now is a brute fact of my personal experience, and  I conclude from what others tell me that it is also a brute fact of the personal experience of everybody I communicate with.  Together we have all deduced, from our direct personal data and the communications of others, an abstract model that we call space-time. The model provides a way for each of us to record our memories, direct and reported, of all these Nows, and our anticipations of subsequent Nows.     

This model singles out no part of space-time as Now.   But to say that Now plays no role in the physical description of space-time is to overlook the crucial fact that my personal Nows constitute the only grounds I have for my physical description of the contents of space-time. This oversight is a fine example of what Schr\"odinger had in mind when he  cited Democritus on the Senses addressing the Intellect:   ``Do you hope to defeat us while from us you borrow your evidence?  Your victory is your defeat.''\ft{{\it Nature and the Greeks and Science and Humanism\/},  Cambridge University
Press (1951).}

Any person's Now is a special event for that person as it is happening.  An {\it event\/} is an experience whose duration and location are constricted enough for it to be useful\ft{Useful to whom?      
Useful to whoever is  constructing the space-time representation.}  to represent it as a point in space and time.   My Now is distinguished from other events I have experienced by being the actual current state of affairs. I can distinguish it from earlier events (former Nows) that I merely remember and from later events that I can only anticipate or imagine.  My remembered past terminates in my Now.  My future comes after it.  The status of any particular event as my Now is fleeting, since it fades into a memory with the emergence of subsequent Nows.

Obvious --- indeed, banal --- as the human content of the preceding paragraph is, such a Now is absent from the conventional physical description of spacetime.   All the events experienced by a person constitute a (time-like) curve in spacetime.  There is nothing about any point that gives it a special status as Now.  But my experience of the Now suggests that my world line ought to terminate in something like a glowing point, signifying my Now. That glow should move in the direction of increasing time, as my world line grows to accommodate more of my experience. There is simply nothing like this in the conventional physical description of my spacetime trajectory.

The problem of the Now will not be solved by discovering new physics behind that missing glowing point. It is solved by identifying the mistake that leads me to conclude, against all my experience, that there is no place for my Now in my existing physical description of the world.   

There are actually two mistakes. The first, as already noted,  lies in the deeply ingrained refusal to admit that whenever I use science, it has a subject (me) as well as an object (my external world).

The second mistake, not unrelated to the first,  is the promotion of spacetime from a useful four-dimensional diagram into what Bohr calls a Òreal essence.Ó My diagram, drawn in any fixed inertial frame, enables me to represent events from my past experience, together with my possible conjectures, deductions, or expectations for events that are not in my past or that escaped my direct attention. By identifying my diagram with an objective reality, I fool myself into regarding the diagram as an objective four-dimensional arena in which my life is lived.\ft{Compare the apocryphal Einstein remark  4.4.3 to Ehrenfest.    The Einstein who reportedly spoke to Ehrenfest is not the Einstein described by Carnap.}   

The events we experience are complex, extended entities, and the clocks we use to locate our experiences in time are extended macroscopic devices. To represent our actual experiences as a collection of mathematical points in a continuous spacetime is a brilliant strategic simplification, but we ought not to confuse a cartoon that attempts concisely to represent aspects of our experience with the experience itself.

If I take my Now as the personal reality it clearly is for me, and if I recognize that spacetime is a diagram that {\it I\/} use to represent {\it my\/} experience, then the problem of the Now disappears. At any moment I can plot my past experience in my diagram as a continuous time-like curve that terminates in the Now. As my Now recedes into memory it ceases to be the real state of affairs and is replaced in my updated diagrams by subsequent Nows.   Each updating is Now for me only at the unique (to within the temporal width --- many milliseconds ---  of my personal Now) moment that it takes place.   

The motion of my Now along my trajectory in my diagram reflects the simple fact that as my wristwatch advances, I acquire more experiences to record in my diagrams. My Now advances at one second of personal experience for each second that passes on my watch. According to special relativity, this means a second of personal experience for each second of proper time along my trajectory. This connection between my ongoing experience and a geometric feature of my diagram is just my diagrammatic representation of the fact that if I am asked what time it is now [Now], I look at my watch and report what I read.

\vskip 15pt 

There is thus no problem of the Now for any single person.  But is there a problem in combining the Nows of two different people? Can we account physically for another obvious primitive fact of human experience.  Whenever two people are together at a single event, then if that event happens to be Now for one of them, then it must be Now for them both. When we are interacting face-to-face, it is simply unimaginable that a live encounter for me could be only a memory for you, or vice versa.  If we believed that,  we could not function as social beings.   But can physics have anything to say about this firm human belief?

This commonality of my Now and your Now whenever we are together requires that our Nows must coincide at each of two consecutive meetings, as indeed we find whenever any of us move apart and then come back together. But at the slow relative speeds at which we invariably move, the possibly complicating effect of relativistic time dilation on the advances of our individual Nows is utterly negligible compared with the psychological width --- a great many milliseconds --- of each of our private Now experiences. Must our Nows still coincide when we come back together, even if we move back and forth at relative speeds comparable to the speed of light?

This is a {\it physics\/} question about the Now.    And physics can answer it.

Physics tells us (without bringing the subject into the story) that if two identical watches, initially together and showing the same time, move apart and then come back together then the readings of the two watches will, in general, no longer agree. If the relative speed of the two watches has been comparable to the speed of light, then the disagreement in their readings can be substantial.   

But (bringing the subject into the story)  {\it my\/} Now advances along {\it my\/} trajectory at one second of {\it my\/} personal experience for each second that passes on {\it my\/} watch, which follows the same trajectory as {\it  I\/} do.  And {\it your\/} Now advances along {\it your\/} trajectory at one second of {\it your\/} personal experience for each second that passes on {\it your\/} watch, which follows the same trajectory as {\it you\/} do.   

Suppose we move apart and back together at such speeds that 3 minutes pass on my watch and 5 minutes on yours, and suppose we each update our diagrams at our first meeting and every minute after that, according to our own watch.  Then as we meet again, I will be doing my third updating since we left each other and you will be doing your fifth.     So my personal experience of my third updating includes my experiencing your doing your fifth updating, and vice-versa.   Our Nows continue to coincide.

Even if our watches have changed by significantly different amounts between the two intersections of our trajectories that represent our two meetings, if our Nows coincide at our first meeting, they will also  coincide at our second meeting.\ft{This is true whether or not each of us chooses to represent all the relevant events in our own private diagram, or we agree to represent them all in a single common diagram.}

So, far from having nothing to say about the Now, physics actually describes it in a way that is consistent with our psychological and social experience of each other, even in a world of  people moving about at relativistic speeds.\ft{A skeptical referee asked what it would be like if our Nows were not in synchronization.    The only difference, if we somehow knew that they were not, is that each of us would know that they were  interacting with a mindless automaton --- the technical term here seems to be ``zombie''.   The point of the argument above is only that if you adopt a CBist perspective then physics itself is fully capable of accounting for the naive human experience of the Now, contrary to Einstein's assertion.} 

But to apply physics to the Now to reach this conclusion, one must  acknowledge that the subject --- the particular user of physics --- is as much a part of the story as the object --- the world external to that user.   Physics does have things to say about that glowing point.   But only if you are willing to acknowledge the relevance of that which the point represents: the personal experience of the user of physics. 

\bigskip

I have been offered the following {\it reductio ad absurdum\/} of the problem of the Now:  How can there be a problem of the Now, when everybody agrees that there is no problem of the Here? 

But everybody should not agree. There {\it is\/} a problem of the Here.    And its solution is immediate, given the QBist/CBist solution to the problem of the Now:   {My Here is where I am Now.\/}    

If you insist on a physics that makes no reference whatever to the individual user of physics, then there is indeed no room in that physics for either the Here or the Now. Making a mystery out of the obvious difference between the present moment and what is past or yet to be, seems an excessive price to pay for maintaining the centuries-old exclusion of the user from the story told by physics.   

So it is not just our understanding of quantum mechanics that has been obscured by the exclusion of the user.  To understand 
the prevailing confusion about quantum mechanics, it helps to understand that the historic exclusion of the user, while it may have helped get science off the ground,  is  today responsible not only for the confusion at the foundations of quantum mechanics, but also for a broader misunderstanding of the nature of science in general.

\vskip 20pt

I first write about my understanding of QBism in  {\it Physics Today.\/}\ft{July, 2012.   Reprinted as Chapter 31 of {\it Why Quark Rhymes with Pork\/}, Cambridge University Press, 2016.}  That brief essay elicited several critical letters to the editor.  They all had one thing in common.  Each writer had no problem understanding what quantum mechanics was all about.   Each described their own understanding --- they were all different --- and, to my disappointment,  had nothing whatever to say either for or against the point of view that I was advocating.   

Since writing that essay I had come to realize that an earlier essay I had published in {\it Physics Today\/}\ft{May, 2009.  Reprinted as Chapter 30 of {\it Why Quark Rhymes with Pork\/}.}  in 2009,  as one of the last of their series of ``Reference Frame" columns, not only foreshadowed  my 2012 exposition of QBism, but described it in a broader setting that applied equally well to classical physics.   It occurred to me that nobody had a personal stake in their favorite interpretation of classical physics.   
Therefore if I published a further essay applying QBist thinking to a problem in strictly classical physics, certified as a profound puzzle by Einstein himself, then it might elicit some genuine criticism, rather than just expositions of the letter-writer's preferred way of solving the problem.    I did so in {\it Physics Today\/},  March 2014,\ft{Reprinted as Chapter 32 of {\it Why Quark Rhymes with Pork.\/}}  To my further disappointment, the article elicited no letters whatever, critical or favorable.  Nor did a companion article in {\it Nature\/}.\ft{{\it Nature\/}, March 26, 2014.} Physicists have no interest in the interpretation of classical mechanics.   That's part of the problem.

\vskip 15pt
\centerline{\titlesm VI.  A few final reflections.}
\nobreak\vskip15pt


The language of science scrupulously avoids mentioning the subject --- the user of science.   So does much of the ordinary language we have learned to use to talk about the external world.    The puzzles and paradoxes of quantum mechanics arise from such omissions.  Putting the scientist back into the story requires us to expand into human terms  all the impersonal constructions that we have become used to and generally rely upon to make concise speech possible.\ft{Cf.~John Bell's remark 4.1.2 about the term ``Professor Peierls.''}   Abstractions like quantum states, waves, particles, trajectories, spatial and temporal locations, energy levels, etc.~are all common impersonal concepts that we have created to help us make sense of our own particular personal experiences.   For most purposes we can safely regard these abstractions as  entities that are properties of a real objective world.  Indeed, we must, if we are not to be swept away by a tidal wave of complex verbal expansions.   In a similar way, it is simpler and more natural to regard space and time as conditions in which we live, rather than, more accurately, as  modes in which we think.   

QBism is needed  to address metaphysical issues lying outside the bounds of practical physical inquiry.\ft{Not all QBists agree with this.   The hope has been expressed that by reformulating quantum mechanics explicitly in terms of probabilistic expectation in the QBist sense,  one might achieve not only a deeper understanding of the theory, but even some clues about its refinement or generalization.}  For practical purposes it doesn't matter if, like most physicists, I confer objective reality on the theoretical abstractions that enable me to calculate the likelihood of my subsequent experience.  But for resolving certain conceptual puzzles like ``are quantum states real?", or ``the measurement problem'', or ``why is the Now excluded from orthodox physics'', it is essential not to reify what are fundamentally intellectual tools, and not to treat what is fundamentally subjective and personal as if it were objective and universal.   
 
Why have we kept fooling ourselves  about such metaphysical issues for so many years?   Speaking only for myself,  I can trace it back to the sentiment that led me in 1956 to graduate school in physics, rather than to law school.   The problem with the law, I felt, was that it was based on principles that  grew out of human conditions --- principles that referred to human judgments or perceptions.   I wanted to devote my life to something more objective and impersonal, less contingent on human imperfection and imprecision.   It can be hard to acknowledge  that it's humanity all the way down, in all fields --- even physical science.   There was no need to acknowledge it until quantum mechanics refused to make sense\ft{Unless, like Einstein, you too were troubled by the problem of the Now.}     
 for almost a century.
 
In 2017 I was asked to contribute my thoughts on these and  related issues to an annual series of essay collections published only in Czech by the V\'aclav and Dagmar Havel foundation.\ft{N. David Mermin, {\it Mysl, Smysl, Sv\u et}, 
Mavlovan\'y Kraj, Praha, 2017.}    Karl Pribram had begun the series in 1999 with {\it Mozek a mysl\/} --- ``brain and mind''. Umberto Eco followed him with {\it Mysl a smysl\/} --- ``mind and meaning''. And Zden\u ek Neubauer picked up the baton and contributed {\it Smysl a sv\u et\/} ---  ``meaning and world''.  In the fifteen titles that followed those first three, the pattern was abandoned.  But Pribram, Eco, and Neubauer  gave me my title for volume nineteen: {\it Mysl, smysl, sv\u et.\/} 

Science starts with {\it mind\/}, 
the private library of experience for each of us.   From the contents of our own experience each of us strives to assemble what that experience {\it means\/} about the
{\it world\/} that gives rise to it. An all too common misreading of QBism in the
popular scientific press is ``It's all mind". This is as wrong as the opinion most physicists  have about
physics, that it's all world. There is mind and there is a world. Quantum mechanics has
taught us that we cannot understand what we are talking about without paying attention to both. What links the contents of my mind to the world that induces them is the meaning I construct
for my experience.  

If I had to design  a coat of arms for QBism, it would display three words: {\it mysl, smysl, sv\u et.\/}   They would have to be in Czech. ``Mind, meaning, world'' has no poetry in it. And what physicists' understanding of quantum mechanics has lacked for ninety years is any hint of poetry.
 
 Paul Dirac is said to have told Robert Oppenheimer that science takes something nobody
can understand and says it in a way that anybody can understand, whereas poetry
takes something everybody can understand and says it in a way that nobody can understand.
In 2017 QBism takes the scientific process, which almost all scientists think they
understand, and states it in a way that only half a dozen currently understand.
It's my hope that this essay may induce a few more people to think about joining us.

\vskip 20pt


\noindent {\it  Acknowledgment.\/}  \hskip 5pt  
I am grateful to Carl Caves, R\"udiger Schack, and, especially, Chris Fuchs, for patiently (most of the time) trying to answer my skeptical questions, reservations, and doubts over the past two decades.    I am indebted to 
Hendrik Geyer for making it possible for me to spend a month and a half with Fuchs and Schack at the Stellenbosch Institute for Advanced Study in 2012, where I finally came to understand what they had been trying for so long to explain to me.    This paper owes its existence to Gordon Baym, who  has  repeatedly urged me, at least since 2009, to write something --- anything --- about my own view of quantum foundational issues.   This final form of my manuscript has benefited  from the close critical reading of an earlier version by a perceptive referee.

\bye